\documentclass{kapproc} % Computer Modern font calls

\usepackage{t1enc}

\usepackage{procps} 

\usepackage[dvips]{graphicx}

\upperandlowercase

\kluwerbib % will produce this kind of bibliography entry:

\begin{document}

\articletitle{Local Starbursts in a Cosmological Context}

\author{Timothy M. Heckman}
\affil{Center for Astrophysical Sciences\\
Johns Hopkins University}
\email{heckman@pha.jhu.edu}

\begin{abstract}
In this contribution I introduce some of the major issues that 
motivate the conference, with an emphasis on how starbursts fit
into the ``big picture''.
I begin by defining starbursts in several different ways, and discuss
the merits and limitations of these definitions. I will argue that
the most physically useful definition of a starburst is its ``intensity''
(star formation rate per unit area). This is the most natural parameter
to compare local starbursts with physically similar galaxies at
high redshift, and indeed I will argue that local starbursts are
unique laboratories to study the processes at work in the early
universe. I will describe how NASA's GALEX mission has uncovered
a rare population of close analogs to Lyman Break Galaxies in the 
local universe. 
I will then compare local starbursts to the 
Lyman-Break and sub-mm galaxies high redshift populations, and speculate 
that the multidimensional ``manifold'' of starbursts near and far can be
understood largely in terms of the Schmidt/Kennicutt law and galaxy
mass-metallicity relation.
I will briefly summarize
the properties of starburst-driven galactic superwinds and their
possible implications for the evolution of galaxies and the inter-galactic
medium. These complex multiphase flows are best studied in nearby starbursts,
where we can study the the hot X-ray gas that contains the bulk of the
energy and newly produced metals.
\end{abstract}


\section{Introduction: What is a Starburst?}

Why are local ($z$ << 1) starbursts important? First of all, they are
a very significant component of our present-day universe, and as such
deserve to be understood in their own right. They provide roughly
10\% of the radiant energy production and about 20\% of all the
high mass star formation in the local universe (e.g. Heckman 1998;
Brinchmann et al. 2004). Their cosmological relevance
has been highlighted by their many similarities to
star forming galaxies at high-redshift. In particular,
local UV-bright starbursts appear to be good analogs to the
Lyman Break Galaxies (Meurer et al. 1997; Shapley et al. 2003;
Heckman et al. 2004). Local starbursts provide
a laboratory in which to study
the complex ecosystem of stars, gas, black holes, galaxies, and the
intergalactic medium up close and in detail. Finally,
starbursts can contain millions of OB stars, and hence they
also offer a unique opportunity to test theories
of the evolution of massive stars.

Perhaps the most fundamental definition of a starburst would be that
it is a galaxy in which the star formation rate approaches the
upper limit set by causality. For a self-gravitating system this upper
limit is reached if the entire gas reservoir is converted into stars in one
dynamical time. For a total mass $M_{tot}$, a gass-mass fraction
$f_{gas}$, and a velocity dispersion $\sigma$,
this upper bound can be written as

$SFR \leq M_{tot}~f_{gas}/t_{dyn} \sim f_{gas}~\sigma^3/G
\sim 115 (\sigma/100 km s^{-1})^3 f_{gas}~M_{\odot}/yr$

If the star formation occurs with a standard IMF, the
implied bolometric luminosity is:

$L_{max} \sim 10^{12} (\sigma/100 km s^{-1})^3 f_{gas}~L_{\odot}$

Figure 1 compares this upper limit to observations of starburst galaxies
in both the local universe and at high redshift. Starbursts approach the
limit, and in extreme cases consistency requires gas-mass fractions
approaching unity.

\begin{figure}[!ht]
\includegraphics[width=4.5in]{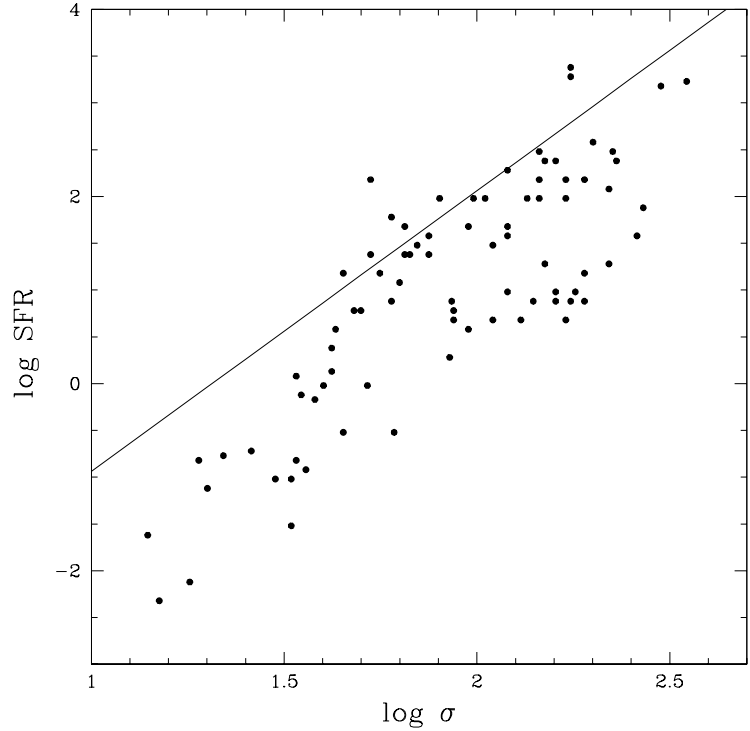}
\caption{The log of the star formation rate ($M_{\odot}$ per year) is
plotted vs. the log of the galaxy velocity dispersion (km/sec) for
the sample of local and high-redshift starbursts described
by Murray et al. (2004). The diagonal line corresponds to the upper bound on
the star formation rate allowed by causuality for a gas mass
fraction of 100\% (it corresponds to the conversion of the entire mass
of the system into stars in a single dynamical time).
See Heckman (1994) for an older version of this figure.
Note that Murray et al. interpret this in terms of
an upper limit on star formation set by the Eddington limit
for radiation pressure acting on dust.}
\end{figure}

The classic definition of a starburst is in terms of its duration, and
has two variations.
First, a starburst is commonly defined as a galaxy in which the time
it would take at the current star formation rate to consume the
remaining reservoir of interstellar gas is much less than the age of
the universe.
This is commonly called the gas consumption time. The inverse of the
gas consumption time is sometimes called the ``efficiency''.
The related definition
is that a starburst is a galaxy in which the time it would take
to produce the current stellar mass at the current star formation rate
is much less than a Hubble time. The inverse of this time can be recast as
the ``birthrate parameter'' ($b$, the ratio of the current to past
average star formation rate).

These are sensible definitions and can
be measured relatively easily. However, it important to note that
the mass ratio of gas and stars varies significantly
and systematically as a function of galaxy properties (e.g. Boselli et al. 
2001).
Thus (for example) using the birthrate parameter as the definition,
leads to a steep decline in the fraction of starbursts with increasing
galaxy mass (e.g. Brinchmann et al. 2004). On the other hand, since
the gas mass fraction is so much smaller in massive galaxies,
using the gas consumption time to define a starburst 
leads to very little mass-dependence of the starburst phenomenon. 
An additional point is that either of these timescale definitions
of a starburst build in a strong redshift-dependence: since 
the age of the universe at $z \sim$ 5 is only 10\% of its present
value, a galaxy with $b$ = 10 (strong starburst) today would have
$b$ = 1 (not a starburst) at $z \sim$ 5.

An alternative definition
is that a starburst has a high intensity:
the star formation rate per unit area ($I_{SF}$) is very large compared
to normal galaxies. As shown by Kennicutt (1998), this definition
is functionally equivalent to the classic definition in terms of gas
depletion time. Since he found that $I_{SF} \propto N_{gas}^{3/2}$,
this means gas depletion time $\propto I_{SF}^{-1/3}$. 
Extreme starbursts have star formation rates per unit area
thousands of times larger than the disk of the Milky Way,
and gas consumption times of only $\sim10^8$ years. I will argue
below that defining starbursts in terms of their intensity
is the most physically useful way to proceed. 

\section{The Consequences of High Intensity}

The very high star formation rate per unit area in starbursts has immediate 
consequences for the basic physical properties of the galaxy.
It immediately implies a high interstellar gas surface mass density
($\Sigma_g$)
(Kennicutt 1998) and also a high stellar surface mass
density ($\Sigma_*$). A typical case would have
$I_{SF} \sim$ 10 M$_{\odot}$ year$^{-1}$ kpc$^{-2}$ and
$\Sigma_g \sim  \Sigma_* \sim$ 10$^9$ M$_{\odot}$ kpc$^{-2}$.
These
are roughly 10$^3$ ($\Sigma_{SFR}$), 10$^2$ ($\Sigma_g$) and
10$^1$ ($\Sigma_*$) times larger than the
corresponding values in the disks of normal galaxies.

The basic physical and dynamical properties of starbursts
follow directly
from the above. A gas surface-mass-density of 10$^9$ M$_{\odot}$ kpc$^{-2}$
corresponds to an extinction of $A_B \sim$ 10$^2$ for a Milky Way
dust-to-gas ratio. The characteristic dynamical time in the star-forming region
will be short: $t_{dyn}$
$\sim$ $(G\rho)^{-1/2}$ $\sim$ $(G\Sigma_{tot}H)^{-1/2}$
$\sim$ few Myr, where $H \sim$ 10$^2$ pc is the thickness of the disk.
A surface-brightness of a few $\times$ 10$^{10}$ L$_{\odot}$ kpc$^{-2}$
corresponds to a radiant energy density inside the star-forming region that
is roughly 10$^3$ times the value in the ISM of the Milky Way. Finally,
simple considerations of hydrostatic equilibrium imply correspondingly
high total pressures in the ISM: $P \sim G \Sigma_g
\Sigma_{tot} \sim$ few $\times$ 10$^{-9}$ dyne cm$^{-2}$ (P/k $\sim$
few $\times$ 10$^7$ K cm$^{-3}$, or several thousand times the value
in the local ISM in the Milky Way). The rate of mechanical energy deposition
(supernova heating) per unit volume is also 10$^3$ or 10$^4$ times
higher than in the ISM of our Galaxy.

As shown by Meurer  et al. (1997), local starbursts and Lyman Break Galaxies
have very similar values for $I_{SF}$ (1 to 100 $M_{\odot}$ yr$^{-1}$
kpc$^{-2}$). 
Thus, this
immediately implies that there are also strong similarities in
the basic physical properties of local starbursts and Lyman Break Galaxies.

\section{Lyman Break Galaxy Analogs at Low Redshift}

While Meurer et al. (1997) showed that local starbursts and
Lyman Break Galaxies have similar UV surface brightnesses,
the former are generally smaller and less luminous than
the latter. Local starbursts are either dwarf galaxies or small
regions (usually nuclear) inside big galaxies.
Are there true local analogs of the Lyman Break Galaxies
in terms of size and ultraviolet luminosity?
The success of NASA's Galaxy Evolution Explorer (GALEX - 
Martin et al. 2004) mission
now makes it possible to find out.
  
We (Heckman et al. 2004) have used the first matched set of GALEX and 
Sloan Digital Sky Survey (SDSS)
data to investigate the properties of a sample of
74 nearby ($z < $ 0.3)
galaxies with far-ultraviolet luminosities greater than
$2 \times 10^{10} L_{\odot}$ (with no extimction correction).
This was chosen to overlap the luminosity range
of typical Lyman Break Galaxies.
GALEX deep surveys have shown that ultraviolet-luminous galaxies
similar to these are the fastest evolving
component of the UV galaxy population 
(Arnouts et al. 2004; Schiminovich et al. 2004).
Model fits to the combined
GALEX and SDSS photometry yield typical FUV extinctions in these galaxies of
0.5 to 2 magnitudes
(similar to Lyman Break Galaxies).
The implied star formation rates are SFR $\sim$ 3 to 30 $M_{\odot}$/year.
This overlaps the range for
Lyman Break Galaxies.
We found a strong inverse correlation between galaxy mass and
far-ultraviolet
surface brightness, and on this basis divided our sample into
``large'' and ``compact'' systems.
The large  ultraviolet-luminous galaxies are
relatively massive ($M_* \sim 10^{11} M_{\odot}$) late type disk galaxies
forming stars at a rate similar to their past average
($M_*/SFR \sim t_{Hubble}$). They are metal rich ($\sim$
solar), and have
intermediate optical-UV colors ($FUV-r \sim$ 2 to 3).
In contrast, the compact ultraviolet-luminous galaxies
 have half-light radii of a
few kpc or less (similar to Lyman Break Galaxies).
They are relatively low-mass galaxies
($M_* \sim 10^{10} M_{\odot}$) with typical velocity dispersions of
60 to 150 km/s.
They span a range in metallicity from $\sim$ 0.3 to 1 times solar, have
blue optical-UV colors ($FUV-r \sim$ 0.5 to 2), and are forming stars at a rate
sufficient to build the present galaxy in of-order a Gigayear. In all these
respects they appear similar to the Lyman Break Galaxies. 

The GALEX mission will ultimately find over a thousand such ``living fossils''
This will provide an opportunity for detailed local
investigation of the physical processes occurring in typical star forming
galaxies in the early universe.

\section{Understanding the Starburst Manifold}

Let us adopt the idea that a starburst by definition has a high
star formation rate per unit area and a short gas depletion time.
The most fundamental properties of a starburst would then be its
star formation rate, metallicity, and dust opacity, and the mass
of its host galaxy. In principle, starbursts could uniformly populate
the multi-dimensional manifold defined by these parameters. Instead,
them parameters show very strong systematic relations. More powerful local
starbursts (higher SFR) are more metal rich, are more dust obscured,
and occur in more massive galaxies (e.g. Heckman et al. 1998). 
At high redshift we know that the highest SFRs also occur in the
most dust-obscured galaxies (the sub-mm sources compared to the
Lyman Break Galaxies).

This systematic behavior can be understood in simple terms as
the consequences of three effects. First, 
as discussed
in the introduction above, 
causality implies that the maximum
possible SFR is higher in more massive galaxies because they have
higher velocity dispersions. This effect is mitigated somewhat
by the systematically lower gas mass fraction ($f_{gas}$) in more
massive galaxies. Second, more massive star forming galaxies have
systematically higher metallicity (Tremonti et al. 2004). Assuming
a constant dust/metals ratio this implies that more massive
galaxies have higher dust/gas ratios in the ISM. Third, the
Schmidt-Kennicutt law implies that
a high SFR requires a high ISM column density.

The natural assumption is that the amount of extinction in a starburst will be
strongly related to the dust column density in the ISM (gas column
density times dust/gas ratio).  
These three effects then natually explain why the more powerful starbursts
are the more dust obscured (Wang \& Heckman 1996; Martin et al. 2004b).
A higher SFR requires both a higher gas column and a more massive galaxy  
with its associated higher dust/gas ratio.

The argument can be easily extended to explain why galaxies with
a given SFR are less dust-obscured at high redshift than in the local
universe (Adelberger
\& Steidel 2000). Consider a galaxy with a specified mass (stars plus gas
plus dark matter).
The Schmidt-Kennicutt law implies that
the specific star-formation rate is $SFR/M \propto f_{gas} N_{gas}^{1/2}$.
As long as $f_{gas}$ is systematically higher at higher redshift,
this then implies that a correspondingly smaller $N_{gas}$ is required at
higher redshift to sustain the same $SFR/M$ (e.g. if $f_{gas}$ increases
by three, $N_{gas}$ would decrease by nine). If the mass-metallicity
relation is displaced to lower metallicity (dust/gas ratio) at higher redshift 
this would only reinforce
the effect. 

The arguments above imply that both rest-frame UV and far-IR data 
are required to get a complete picture of the population of star forming
galaxies, both locally at at high redshift.
A UV(FIR)-selected sample
will preferentially sample the population with lower (higher)
metallicity, SFR, and mass
(Martin et al. 2004b; Buat et al. 2004).
In light of the above, it is natural to speculate that the main
physical difference
between the population of star forming galaxies at high-z selected 
as Lyman Break Galaxies or sub-mm sources is the galaxy mass. The latter
may well be the progenitors of giant elliptical galaxies.

\section{Starburst-Driven Galactic Winds}

By now, it is well-established that galactic-scale outflows of gas
are a ubiquitous phenomenon in the most actively star-
forming galaxies in the local universe (see Heckman 2002 for a recent review).
These outflows are potentially very important in the evolution of galaxies and
the intergalactic medium. For example, by selectively blowing metals out of
shallow galactic potential wells, they may explain the tight relation 
between the mass and metallicity in galaxies
(Larson 1974; Tremonti et al. 2004). This same process
would have enriched and heated the intergalactic medium in metals at early
times (e.g. Adelberger et al. 2003),
and explain why the majority of metals in galaxy clusters are in the
intracluster medium (e.g. Loewenstein 2004). 

We know that galactic winds are ubiquitous in the population
of Lyman Break Galaxies (e.g. Shapley et al. 2003), and have also
been observed in sub-mm-selected galaxies (Smail et al. 2003). The
big advantage in studying them in local starbursts is that we
can investigate their physics in considerably more detail, and
in so-doing better understand the form and magnitude of the mass,
energy, and metals being carried out in the wind.

%\begin{figure}[!ht]
%\includegraphics[width=4.5in]{f2.eps}
%\caption{The blueshift of the centroid of the interstellar absorption
%lines (relative
%to the galaxy systemic velocity) is plotted as a function of the starburst
%luminosity (taken as the sum of the far-UV and far-IR luminosity) for a
%sample of 21 starbursts observed by FUSE. Outflows at velocities of
%over a hundred km/s are common in powerful starbursts. The seven data points
%plotted at log $v_{out}$ = 1.2 represent upper limits on outflow speed.}
%\end{figure}

Observations of local starburst winds show that they are complex
multiphase phenomena. The hot ($\sim10^7$ K) gas traced by X-ray
emission appears to arise in shocks in the wind fluid as it impacts
cooler denser material in the galaxy halo (e.g. Strickland et al. 2004;
Lehnert, Heckman, \& Weaver 1999). As this ambient material
encounters the wind it is heated and accelerated, giving rise
to regions of optical line emission and the 
blueshifted interstellar absorption lines that are characteristic
of Lyman Break Galaxies (Shapley et al. 2003) and local starbursts
(e.g. Heckman et al. 2000). Dust contained in these clouds is revealed
as it reddens the background starlight (Heckman et al 2000) and
scatters the starburst's UV radiation (Hoopes et al. 2004).

The picture emerging from these panchromatic investigations
of local starburst winds is that the fate of the outflow
depends strongly on the phase of the outflow and the mass
of the galaxy. The hot gas (which
contains most of the energy and metals) has nearly the same
temperature, independent of the escape velocity from the galaxy blowing the
wind (Martin 1999).
This hot gas
is thus more likely to escape from low mass galaxies (with
their shallower potential wells). This could naturally account
for the galaxy mass-metallicity relation. In contrast, the outflow
velocity in the cooler gas traced by the interstellar absorption lines
is lower in the much less powerful starbursts in dwarf galaxies
(Martin 2004). This is most likely
because the low power winds in the dwarfs have insufficient 
thrust to accelerate interstellar clouds up to the velocity
of the hot wind.

The combination of insights like these from the local universe
and systematic investigations
of the redshift dependence of the outflow rate derived through
rest-frame UV spectroscopy will prove quite powerful in terms
of addressing the cosmological significance of starburst-driven winds.

\section{Conclusions}

\begin{itemize}
\item
Starbursts are an important component of the local universe and worth
understanding in their own right. 
\item
The key astrophysical property of a starburst is its
``intensity'' (SFR/area). This property has far-reaching consequences
for the physical and dynamical properties of the ISM.
\item
Local starbursts provide excellent laboratories
for the study of the astrophysics of high-z starforming galaxies, which have
very similar SFR/area.
\item
GALEX has begun to provide a large sample of low-z analogs to Lyman Break
Galaxies.
\item
The systematic properties of starbursts are largely a consequence
of the Schmidt/Kennicutt law plus the mass-metallicity relation. This explains
why more massive galaxies host more powerful, more metal-rich, and
more highly obscured starbursts. It also explains why high-z starbursts
are less obscured on-average for a given SFR than low-z starbursts.
\item
Only in local starbursts can the multi-phase physics of galactic
winds be fully investigated. The hot metal-rich phase traced by
X-rays is the key to understanding how the IGM was chemically enriched
by outflows from low mass galaxies.
\end{itemize}

%\begin{figure}[ht]
%\vskip.2in
%\caption{Short caption.}
%\end{figure}

\end{document}